\documentclass[prl,aps,twocolumn,showpacs]{revtex4}
\usepackage{amsmath,amssymb,graphicx}
\usepackage{epsfig}
\usepackage{textcomp}

\begin{document}
\title{Negative velocity fluctuations of pulled reaction fronts}
\author{Baruch Meerson}
\affiliation{Racah Institute of Physics, Hebrew University of
Jerusalem, Jerusalem 91904, Israel}
\author{Pavel V. Sasorov}
\affiliation{Institute for Theoretical and Experimental
Physics, Moscow 117218, Russia}

%02.50.Ga   Markov processes
%87.23.Cc   Population dynamics and ecological pattern formation
%05.10.Gg	Stochastic analysis methods (Fokker-Planck, Langevin, etc.)
%87.18.Tt 	Noise in biological systems
\begin{abstract}
The position of a reaction front, propagating into an unstable state, fluctuates because of the shot noise.
What is the probability that the fluctuating front
moves considerably slower than its deterministic counterpart? Can the noise arrest the front motion for some time,
or even make it move in the wrong direction?
We present a WKB theory that assumes many particles in the front region and answers these questions for the microscopic model
$A\rightleftarrows2A$ and random walk.
\end{abstract}
\maketitle

The Fisher-Kolmogorov-Petrovsky-Piscounov (FKPP)
equation \cite{Fisher},
\begin{equation}\label{FKPP}
    \partial_t q = q-q^2+\partial_{x}^2 q\,,
\end{equation}
describes invasion of an unstable phase, $q(x\to \infty,t)=0$, by a stable phase, $q(x\to -\infty,t)=1$.
It is one of the
most fundamental models in mathematical genetics and population biology \cite{Fisher,Murray}, but  similar
equations appear
in chemical kinetics \cite{Douglas},
extreme value statistics \cite{Majumdar}, disordered systems \cite{Spohn}
and even particle physics \cite{particlephys}.

The invasion fronts are described by traveling-front solutions (TFSs) of Eq.~(\ref{FKPP}): $q(x,t)=Q_{0,c}(\xi)$, where $\xi=x-ct$. $Q_{0,c}(\xi)$ solves the  equation
\begin{equation}\label{MFeq}
    Q_{0,c}^{\prime\prime}+c Q_{0,c}^{\prime}+Q_{0,c}-Q_{0,c}^2=0\,,
\end{equation}
where the prime denotes the $\xi$-derivative.  It is well known \cite{Saarloos} that,
for a steep enough initial condition, the invasion front of Eq.~(\ref{FKPP}) converges at long times to the limiting TFS, $Q_{0,2}(\xi)$, of Eq.~(\ref{MFeq}) with the velocity $c_0=2$. This special velocity
is determined by the dynamics of the leading edge of the front, where Eq.~(\ref{FKPP}) can be linearized around $q=0$. One can say that the nonlinear front, described by Eq.~(\ref{FKPP}), is ``pulled" by its leading edge, hence the term ``pulled fronts" \cite{Saarloos}, of which the FKPP equation (\ref{FKPP}) is the celebrated example.

Being a mean-field equation, Eq.~(\ref{FKPP}) ignores discreteness of particles and the resulting shot noise. Their impact on the front propagation is dramatic, and it has attracted a great deal of interest \cite{Breuer,Derridashift,Levine,Derridanum,Doering,Panja1,Panja2,Derridatheory,Hallatschek}.  The noise leads to fluctuations of the front shape \cite{Breuer,Doering}. The particle discreteness and noise also cause deviations of the front position with time that include a systematic part  -- the front velocity shift -- and a fluctuating part. If $N\gg 1$ is the number of particles in the front region, the front velocity shift scales as $\ln^{-2} N$ \cite{Derridashift,Levine,Derridanum}, whereas the front diffusion coefficient scales as $\ln^{-3} N$ \cite{Derridanum,Panja1,Derridatheory}. These properties of fluctuating pulled fronts are markedly different from those of fluctuating fronts propagating into\emph{ metastable} states, where the front velocity shift and the front diffusion coefficient scale as inverse powers of $N$ and are therefore much smaller \cite{Kessler,Panja2,MSK}.

The front diffusion coefficient probes typical, relatively small fluctuations of the front position.  Here we ask the following question that has not been addressed before.  What is the probability ${\cal P}(c)$ that a fluctuating front moves, during a long time interval $\tau$, with average velocity $c$ that is considerably smaller than the mean-field value $c_0=2$? This includes the extreme case of $c=0$, when the front is standing on average,
and even $c<0$, when it moves ``in the wrong direction." We will also obtain new results in the regime of $2-c\ll 1$. Importantly, these results can be extended to a broad class of pulled fronts. Still,  we will focus on a specific microscopic model which mean-field limit is Eq.~(\ref{FKPP}). A WKB (after Wentzel, Kramers and Brillouin)
formalism transforms the governing master equation into an effective Hamiltonian mechanics, continuous in space and time. Then the problem can be reduced to finding a TFS, with given $c$, of the Hamilton's equations. This TFS turns out to be instanton-like, as it connects two ``equilibria."
The action along the instanton yields, up to a pre-factor, ${\cal P}(c)$ which is strongly non-Gaussian,
exponentially small in $N$ and rapidly falls as $c$ goes down.

The microscopic model that we adopt here, see \textit{e.g.} Ref. \cite{Levine,Doering},  involves particles on a one-dimensional lattice with lattice constant $h$. The particles undergo stochastic on-site reactions $A\to 2A$ and $2A\to A$ with rate constants $\lambda$ and $2\lambda/K$, respectively, and unbiased random walk between neighboring sites with rate constant $D_0$. For $K\gg 1$
and $D_0\gg \lambda$, the mean-field theory for this problem is Eq.~(\ref{FKPP}), see \textit{e.g.} Ref.  \cite{MS}. Here $q$ is (the continuous limit of) the on-site number of particles $n_i$ rescaled by $K$, time is rescaled by $1/\lambda$, $x$ is rescaled by
$l_D=(D/\lambda)^{1/2}$, and $D=D_0 h^2$ is the diffusion coefficient. The front velocity is measured in units of $(\lambda D)^{1/2}$.

The stochastic problem is formulated in terms of the master equation for the
multivariate probability distribution $\pi(\mathbf{n}; t)$ \cite{Levine,MSK,MS,Gardiner} which describes how the probability of having
$n_i$ particles on site $i$ at time $t$ changes via each of the elementary processes.
We apply the WKB ansatz
$\pi(\mathbf{n}; t) = \exp[-K S(\mathbf{q}; t)]$, with $\mathbf{q}=\mathbf{n}/K$, to the master equation, see e.g. Ref. \cite{MS}. As $K\gg 1$, we can treat $S$ as a smooth functional that can be Taylor expanded in $\mathbf{q}$. Furthermore, in the fast diffusion regime,  $D_0\gg \lambda$, we can use a continuous description in $x$.
In the leading order in $1/K$, the master equation reduces to a functional Hamilton-Jacobi equation
$\partial_t S+H(\mathbf{q},\partial_\mathbf{q} S)=0$. The Hamiltonian functional $H=h^{-1}\int dx \,w$, where
\begin{equation}\label{w}
    w=q (e^p-1) +q^2 (e^{-p}-1) -\partial_x q \partial_x p+ q (\partial_x p)^2\,.
\end{equation}
The Hamilton's equations for the ``coordinate" $q(x,t)$ (the particle number density) and conjugate
momentum $p(x,t)$ are
\begin{eqnarray}
\partial_t q&=&q e^p - q^2 e^{-p}+ \partial_x^2q -2\partial_x\left(q\partial_x p\right)\,,
\label{p100}\\
\partial_t p&=& -(e^p-1) -2q (e^{-p}-1)
-\partial_x^2 p -\left(\partial_x p\right)^2\,.
\label{p110}
\end{eqnarray}
We assume that the initial particle density is zero to the right of some finite point in space.
At $x\to -\infty$ there is a stationary distribution of the particle density, peaked at $q=1$.
Therefore, we demand $q(-\infty,t)=1$ and
$p(-\infty,t)=0$ which corresponds to the fixed point $(q=1,p=0)$ of the \emph{on-site} Hamiltonian
$H_0(q,p)= q (e^p-1) +q^2 (e^{-p}-1)$. We also demand $q(\infty,t)=0$. The momentum
$p$ can be unbounded at $x=\infty$, but must be bounded at finite $x$.

The calculations greatly simplify in the new
variables
$Q=q e^{-p}$
and $P=e^p-1$ \cite{EK,optimal}. The
generating function of this canonical transformation is $h^{-1}\int dx F(q,Q)$, where
\begin{equation}\label{genf}
    F(q,Q)=q\,\ln (q/Q) -q+Q\,.
\end{equation}
The new Hamiltonian density becomes
\begin{equation}\label{density}
\mathcal{W}=(Q-Q^2)(P+P^2) - \partial_x Q\, \partial_x P\,,
\end{equation}
whereas the Hamilton's equations are
\begin{eqnarray}
  \partial_t Q  &=& Q-Q^2+2Q P
  -2 Q^2 P +\partial_x^2 Q\,, \label{1B}\\
  \partial_t P &=& -P-P^2+2Q P+
  2 Q P^2 -\partial_x^2 P\,. \label{2B}
\end{eqnarray}
The mean-field FKPP equation (\ref{FKPP}) corresponds to the motion on
the invariant manifold $P(x,t)=0$.

The boundary conditions (BCs) at $x=\pm \infty$ are determined by the zero-energy
fixed points $(Q,P)$ of the on-site Hamiltonian $\mathcal{H}_0=(Q-Q^2)(P+P^2)$. There are four such  points: $(1,0)$,
$(0,0)$, $(0,-1)$, and $(1,-1)$.  In view of the BCs for $q$ and $p$ at $x=-\infty$,
we demand $Q(-\infty,t)=1$ and
$ P(-\infty,t)=0$. Now, as $q=Q(P+1)$, the BC
$q(\infty,t)=0$ can be satisfied at any of the three fixed points:
$Q(\infty,t)=0, \,
P(\infty,t)=-1$ (we will call
these BCs ``BCs A"),
$Q(\infty,t)=1,\,P(\infty,t)=-1$ (BCs B), or trivial BCs $Q(\infty,t)=P(\infty,t)=0$.  It turns out that
each of the BCs A and B is
possible for fluctuating fronts, whereas the trivial BCs are only possible for mean-field fronts,
where $P(x,t)\equiv 0$. The BCs \emph{in time} involve kink-like particle density profiles $q_1(x)$ and $q_2(x)$, at $t = 0$ and $t = \tau$, respectively, separated by distance $X=c\tau$. Once the Hamilton's equations are solved, one can calculate the action $\mathcal{S}$ along the whole trajectory $q(x,t),\,p(x,t)$ and evaluate the probability ${\cal P}(c)$:
\begin{eqnarray}
 &&-\ln {\cal P}(c)\simeq K \mathcal{S}  = N \int_{-\infty}^{\infty}\! dx \int_0^{\tau} \! dt  \left[p(x,t) \partial_t q-w\right] \nonumber \\
  && \!=  \!N \int_{-\infty}^{\infty}\! dx \,\big\{ \Delta F+\!\int_0^{\tau} \! dt
  \left[P(x,t) \partial_t Q-\mathcal{W}\right]\big\},
  \label{action}
\end{eqnarray}
where $\Delta F$ is the increment of the generating function density (\ref{genf}) between $t=0$ and $\tau$, and $N=K l_D/h \gg 1$.

Our crucial assumption, see also Ref. \cite{MSK}, is that the probability to observe, during
a sufficiently long time $\tau$, a fluctuating front with a given front position
$X$ is determined by the action calculated along a TFS,  $Q=Q(x-ct)$ and $P=P(x-ct)$ of Eqs.~(\ref{1B}) and (\ref{2B}).
This TFS solves the coupled ordinary differential equations
\begin{eqnarray}
 Q^{\prime\prime}+c Q^{\prime}
 + Q-Q^2+2Q P-2 Q^2 P &=&0\,, \label{ODEQ}\\
 P^{\prime\prime}-c P^{\prime}
 +P+P^2-2Q P-2 Q P^2 &=&0\,. \label{ODEP}
\end{eqnarray}
subject to the proper BCs.  Equations~(\ref{ODEQ}) and (\ref{ODEP}) possess a conservation law: $\mathcal{H}_0[Q(\xi),P(\xi)]+Q^{\prime} P^{\prime}=\mbox{const}$,
where the constant is zero because of the BCs.

For a TFS Eq.~(\ref{action}) can be simplified. As we will see,
$\Delta F=0$ for $c>-2$. In this case we obtain
\begin{equation}\label{KS1}
   -\ln {\cal P}(c)\simeq K \mathcal{S}=N\tau\,\dot{s},
\end{equation}
with rescaled action accumulation rate
\begin{equation}\label{accumrate}
   \dot{s}=-\int_{-\infty}^{\infty} \! d\xi (c P Q^{\prime}+\mathcal{W})= \int_{-\infty}^{\infty} \!  d\xi  \left(Q-Q^2\right) P^2\,.
\end{equation}
We will also see that, for $c\leq -2$,
it is the term $P\partial_t Q-\mathcal{W}$ in Eq.~(\ref{action}) that
vanishes, and the only contribution to $S$ comes from $\Delta F$.

The model $A\rightleftarrows 2A$ and random walk, as described by the WKB theory,
has two remarkable symmetries:

1. Let $Q_1(x,t)=u(x,t)$ and $P_1(x,t)=v(x,t)$ be a solution to Eqs.~(\ref{1B})
and (\ref{2B}) obeying BCs A. Then $Q_2=-v(-x,-t)$ and $P_2=-u(-x,-t)$
is also a solution obeying these BCs. Notice,
that $Q$ and $P$ interchanged. In particular, for TFSs we have
$Q_1=u(x-ct)$ and $P_1=v(x-ct)$, and
$Q_2=-v(-x+ct)$ and $P_2(x,t)=-u(-x+ct)$.
Now, if there is a unique TFS (up to a shift in $\xi$), we
obtain
\begin{equation}\label{symm1}
    P(\xi)=-Q(\xi_0-\xi)\,,
\end{equation}
that is $Q$ and $P$ for the same TFS are related.

2. Let $Q_1(x,t)$
and $P_2(x,t)$ be a solution to Eqs.~(\ref{1B})
and (\ref{2B}) obeying BCs A. Then $Q_2=P_1(x,-t)+1$
and $P_2=Q_1(x,-t)-1$ is also a solution obeying these BCs.
$Q$ and $P$ again interchanged!
This
yields a relation between fluctuating TFSs with velocity $c$
and $-c$:
\begin{eqnarray}
  P_{(-c)}(\xi)&=&  Q_{(c)}(\xi+\mbox{const})-1\,. \label{symm}
\end{eqnarray}
In variables $q$ and $p$ this is simply $q_{(-c)}(\xi)=q_{(c)}(\xi+\mbox{const})$: reversibility stemming from the detailed balance property of the microscopic model.  As a result,
the actions for $c<0$ and $-c>0$ are simply related:
\begin{equation}\label{accumrateminusfinal}
    \dot{s}\big|_{(c)}=\dot{s}\big|_{(-c)}-c\,,
\end{equation}
providing one more example of the Onsager-Machlup relation \cite{Kurchan}.

\textsc{$c=0$: Exact solution.} How rare is the situation when the front motion is \emph{arrested} by noise during a sufficiently long time
$\tau$? For $c=0$ Eq.~(\ref{symm}) reduces to $P(x)=Q(x)-1$, and the solution of Eqs.~(\ref{1B})
and (\ref{2B}) with BCs A   is elementary:
\begin{equation}\label{solQ}
    Q(x)=(1+e^{x})^{-1},\;\;\;\; P(x)=-(1+e^{-x})^{-1}\,,
\end{equation}
where we have arbitrarily fixed the front position.
Here $\Delta F=0$, and we can evaluate the action from Eqs.~(\ref{KS1}) and (\ref{accumrate}).  We obtain $\dot{s}=1/3$,
so $\mathcal{P}(c=0)\sim \exp(-N \tau/3)$. In the variables $q$ and $p$ the solution (\ref{solQ}) becomes
$q(x)=(1+e^x)^{-2}$ and $p(x)=-\ln(1+e^x)$.

\textsc{$c\leq -2$: Exact solution.} Another important observation is that $Q(x,t)=1$ solves the Hamilton's equation  (\ref{1B}) and obeys the BCs B. As $Q=q e^{-p}$, we realize that $p=\ln q$ is an invariant
manifold of
Eqs.~(\ref{p100}) and (\ref{p110}) for $q$ and $p$. Elimination of $p$ reduces each of Eqs.~(\ref{p100}) and (\ref{p110}) to
$\partial_t q =-\left(q-q^2+\partial_{x}^2 q\right)$,
which is a time-reversed Eq.~(\ref{FKPP}).  The legitimate TFSs, with $q(\xi=-\infty)=1$ and $q(\xi=\infty)=0$, are those with $q\geq 0$. Therefore, we arrive at
$q_{(c)}(\xi)=Q_{0,-c}(\xi+\mbox{const})$,
for any $c \in (-\infty, -2]$.

Now we can calculate the action for these fronts, using the first line of Eq.~(\ref{action}). As $p=\ln q$, $w=0$.
The integral over $t$ can be evaluated by parts, giving
$\Delta F \equiv f[q(x,\tau)]-f[q(x,0)]$, where $f(q)=q \ln q -q$. We evaluate  $\Delta F$ on the TFS with a given $c$. The integration over $x$ yields $-c\tau$, and we again arrive at Eq.~(\ref{KS1}), with $\dot{s}=-c$. The same result
follows from the second line of Eq.~(\ref{action}), where the only contribution comes from
$\Delta F$.

The origin of the simplicity of the case of $c\leq -2$ is in
the \emph{equilibrium} properties of the reversible microscopic model.  In the discrete version of the model,
the equilibrium multivariate distribution of the number of particles in a finite system with $N_{max}$ lattice sites, is
the product Poisson distribution  \cite{Doering}: $\mathrm{P}(\mathbf{n})=
\prod\limits_{i=1}^{N_{max}}\frac{K^{n_i}e^{-K}}{n_i!}$.
Assuming $n_i\gg 1$ and using the Stirling's formula, one obtains
$\ln \mathrm{P}(\mathbf{n})\simeq -N V\{q(x)\}$, where $V\{q(x)\}=\int_0^L dx \left(q \ln q -q +1\right) \equiv \int_0^L dx \,F(q,1)$,
$L=N_{max} h$, and we have returned to the continuous descriptions in $q$ and $x$. Now one can
consider, for $L \to \infty$, a TFS moving with velocity $c\leq -2$ and evaluate $\mathcal{P}(c)$
arriving at Eq.~(\ref{KS1}) with $\dot{s}=-c$. Note that, as the function $\dot{s}(c)=-c$ is not convex, TFSs
may not correspond to the most likely particle density profiles for $c\le -2$.

\begin{figure}[ht]
\includegraphics[width=2.2 in,clip=]{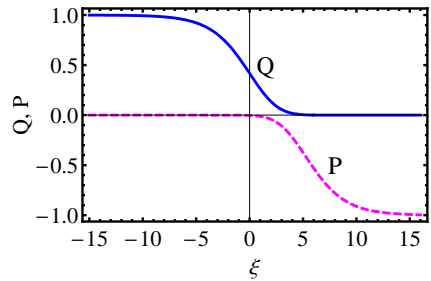}
\caption{Numerical $Q(\xi)$ and $P(\xi)$ profiles for $c=1.5$.}
\label{1.5}
\end{figure}

\begin{figure}[ht]
\includegraphics[width=2.2 in,clip=]{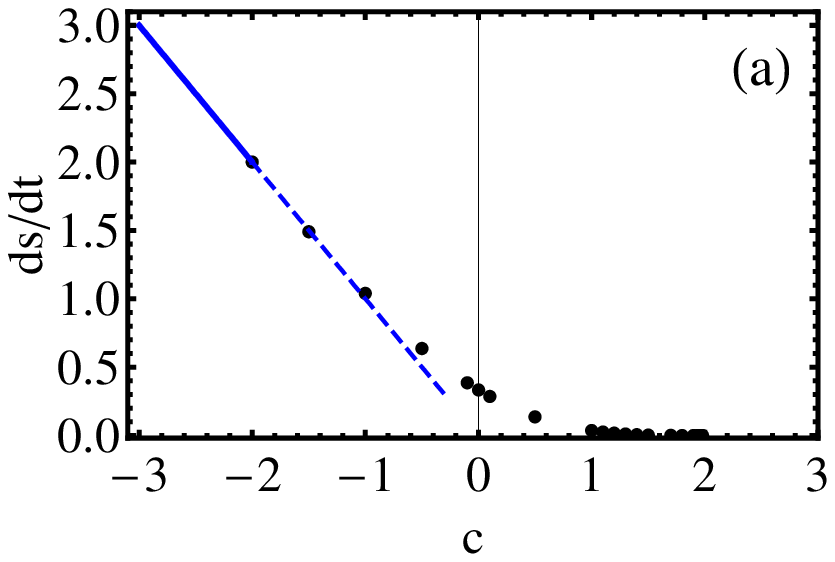}
\includegraphics[width=2.2 in,clip=]{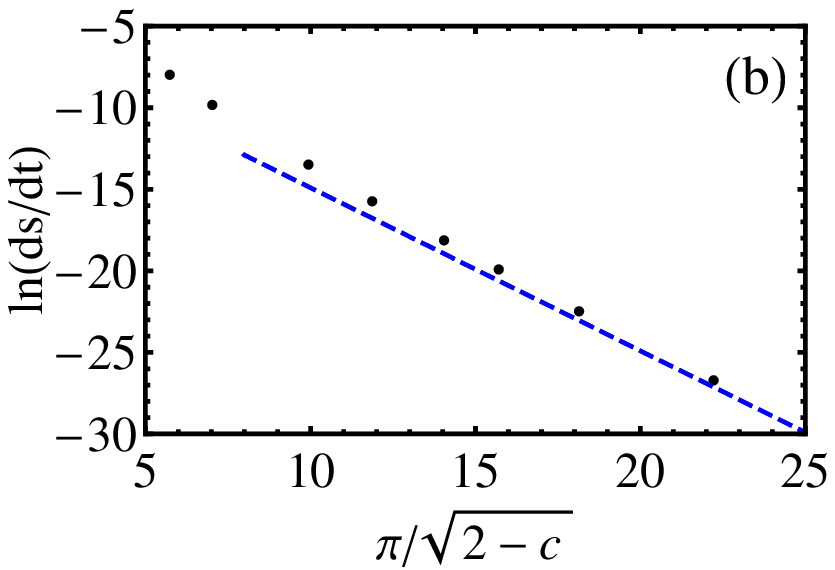}
\caption{(a) Symbols: the numerically found
action accumulation rate $\dot{s}$, see Eq.~(\ref{accumrate}),
versus the front velocity $c$. Straight line: the asymptote
$\dot{s}=-c$ which becomes exact at $c\leq -2$.
(b) $\ln \dot{s}$ versus $\pi/\sqrt{2-c}$ for $c\geq 1.5$. Symbols: numerical results. Dashed
line: asymptote (\ref{asymplead}).}
\label{numaction}
\end{figure}

\textsc{Numerics.} For an arbitrary $-2<c<2$ one needs to solve Eqs.~(\ref{ODEQ}) and (\ref{ODEP}) numerically.
We used a shooting
algorithm described elsewhere \cite{MSK}. In view of Eqs.~(\ref{symm}) and (\ref{accumrateminusfinal})
it would suffice to find the numerical solutions for $0<c<2$. Still, we computed several cases of $c<0$, and verified this symmetry.
One example of numerically found $Q$ and $P$ is shown in Fig. \ref{1.5}. Figure~\ref{numaction}a shows the numerically evaluated $\dot{s}$, see Eq.~(\ref{accumrate}), at different $c$.

One can notice from Fig. \ref{1.5} that the $P$-profile  is shifted ``downstream" compared with the $Q$-profile.  The shift
is equal to zero for $c=0$, see Eq. (\ref{solQ}), and increases with $c$. As $c$ approaches $2$, the shift becomes very large. That is,
$Q(\xi)$ first goes down to zero,
while $P\simeq 0$ remains almost unchanged. Then $P$ goes down
to $-1$ while $Q$ is already close to zero. This salient feature calls for a perturbation theory that we will now present.

\textsc{$2-c \ll 1$: Nonlinear perturbation theory}. At $2-c \ll 1$ the main contribution to the
action comes from the leading edge of the front, $\xi\gg 1$, where $Q\ll 1$.
As $|P| \sim 1$ there, this theory
must be nonlinear. This is in contrast to the front propagation into a metastable state,
where a \emph{linear} theory holds for $c$ close to its mean-field value  \cite{MSK}.
As one can see from the numerical solutions, $|P|\ll 1$ in the left region,
whereas $Q\ll 1$ in the right region. An analytical theory is possible
because of the presence of a \emph{joint} region, where $|P|\ll 1$ and $Q\ll 1$ simultaneously.

In the left region we can drop terms with $P$ in Eq.~(\ref{ODEQ}) and approximate
$Q(\xi)$ by the mean-field solution $Q_{0,c}(\xi)$, where $2-c \ll 1$. Now we linearize Eq.~(\ref{ODEP}):
$P^{\prime\prime}-c P^{\prime}
 +P-2Q_{0,c}(\xi) P=0$.
The solution decaying at $\xi\to -\infty$ is $P_{left}(\xi)=p_0 Q_{0,c}^{\prime}(\xi) e^{c\xi}$,
with a yet unknown constant $p_0\ll 1$. In the right region we can drop terms with $Q$ in Eq.~(\ref{ODEP}):
$P^{\prime\prime}-c P^{\prime}
 +P+P^2=0$. Because of the symmetry (\ref{symm1}), the solution is
\begin{equation}\label{Prightsol}
    P_{right}(\xi)=- Q_{0,c}(\xi_0-\xi)\,,
\end{equation}
with a yet unknown $\xi_0\gg 1$.  To find $p_0$ and $\xi_0$, we match the asymptotes,
$P_{left}(\xi)=P_{right}(\xi)$, in the joint region. Here
both $\xi$ and $\xi_0-\xi$ are much greater than $1$, and we can use
the leading-edge asymptote
$Q_{0,c}(\xi\gg 1)\simeq A (2-c)^{-1/2}\,e^{-\xi}\,\sin\left(\sqrt{2-c} \,\,\xi\right)$,
where $A\sim 1$, and we have fixed the front position so that the cosine term is absent.
The matching yields
\begin{equation}\label{xi0}
     \xi_0  = \pi/\sqrt{2-c} +1 + {\cal O}\left(\sqrt{2-c}\right),\;\;\;\;\;p_0=e^{-\xi_0}.
\end{equation}
Now $P(\xi)$ is known, in terms of the overlapping asymptotes $P_{left}$ and $P_{right}$, for
all $\xi$.
By symmetry,
\begin{equation}\label{Qsol}
Q(\xi)=\left\{\begin{array}{ll}
Q_{0,c}(\xi)\,,  &\mbox{$\xi_0-\xi \gg 1$}, \\
-p_0 Q_{0,c}^{\prime}(\xi_0-\xi) e^{c(\xi_0-\xi)}\,, &\mbox{$\xi\gg 1$}.
\end{array}
\right.
\end{equation}
Now we calculate $\dot{s}$ from (\ref{accumrate}).
The main contribution to the integral comes from the leading edge, $\xi\gg 1$, where
$Q^2$ can be neglected.
Using Eqs.~(\ref{Prightsol}) and (\ref{Qsol}) and keeping the leading and subleading terms
in $\sqrt{2-c}$ in the exponent, we obtain after integration by parts
\begin{equation}\label{asymplead}
   \dot{s}= \frac{2}{3 e}\, e^{-\frac{\pi}{\sqrt{2-c}}}
\int_{-\infty}^{\infty}\!\!
e^{2 \zeta} Q_{0,2}^3(\zeta)\,d\zeta \simeq 0.0074 \, e^{-\frac{\pi}{\sqrt{2-c}}}.
\end{equation}
where $Q_{0,2}(\xi)$ is fixed by the demand
that  $Q_{0,2}(\xi\gg 1)\simeq A \xi e^{-\xi}$, whereas the $ e^{-\xi}$ term is absent.

\textsc{In summary}, we evaluated, for the microscopic model $A\rightleftarrows2A$ and random walk,
the probability $\mathcal{P}(c)$ of observing a fluctuating FKPP front moving with velocity $c<2$: see Eq. (\ref{KS1}) and Fig. \ref{numaction}. For $2-c\ll 1$ we found
\begin{equation}\label{P}
    \ln \mathcal{P}(c) \simeq  -0.0074\, N \tau \exp (-\pi /\sqrt{2-c})\,.
\end{equation}
This result holds only when there are many particles at the leading edge of the front, where $\xi\simeq \xi_0\simeq \pi/\sqrt{2-c}$. This demands $c_{*}-c\gg 2\pi^2 \ln^{-3} N$, where $c_{*}=2-\pi^2 \ln^{-2} N$. Somewhat surprisingly, $c_*$ coincides with the velocity of the mean-field FKPP front with account of discreteness of particles \cite{Derridashift}. In view of the strong inequality $c_{*}-c\gg 2\pi^2 \ln^{-3} N$, Eq.~(\ref{P}) may have a joint
region of validity with the results of Brunet \textit{et al.} \cite{Derridatheory}. This issue
demands a careful study \cite{Dhelp}. We only note here that, by putting $c=c_*-\delta c$,
where $\delta c \ll \ln^{-2} N$, the linear dependence on $N$ in Eq.~(\ref{P}) cancels out,
leading to a much slower logarithmic dependence, \textit{cf.} Ref. \cite{Derridatheory}.

Finally, we have found that Eq.~(\ref{P})
holds, up to a $c$-independent coefficient, for all pulled reaction fronts,
where $A\to 2A$ is the only first-order birth process. It also holds, close to the transition, for all sets of reactions belonging to the Directed Percolation universality class. The properly rescaled on-site Hamiltonian for this class of models is
$H_0(Q,P)=QP(P-Q+1)$ \cite{EK}.

We are very grateful to Bernard Derrida for drawing our interest to fluctuating pulled fronts
and
for illuminating discussions.
This work was supported by the Israel Science Foundation (Grant No.
408/08) and by the Russian Foundation for Basic
Research (Grant No. 10-01-00463).


\begin{thebibliography}{99}
\bibitem{optimal} Note that $q$ rather than $Q$ describes the most likely particle density profile of a noisy front which exhibits a large negative velocity fluctuation.
\end{thebibliography}
\end{document}